# Exploring the development and deployment of a C++ based HPC system for MapReduce as a Hadoop JVM alternative

S. Vignesh
Computer Science Dept.,
*Vellore Institute of Technology, Chennai*
*Tamil Nadu, India*
s.vignesh2017a@vitstudent.ac.in

V. Muthumanikandan
Computer Science Dept.,
*Vellore Institute of Technology, Chennai,*
*Tamil Nadu, India*
Muthumanikandan.v@vit.ac.in

S. Siddarth.
Computer Science Dept.,
*Vellore Institute of Technology, Chennai,*
*Tamil Nadu, India*
siddarthsairaj2017@vitstudent.ac.in

G. Sainath.
Computer Science Dept.,
*Vellore Institute of Technology, Chennai*
*Tamil Nadu, India*
Sainath.g2017@vitstudent.ac.in

*Abstract*— MapReduce is a technique used to vastly improve distributed processing of data and can massively speed up computation. Hadoop and its MapReduce relies on JVM and Java which is expensive on memory. High Performance Computing based MapReduce framework could be used that can perform more memory-efficiently and faster than the standard MapReduce. This paper explores an entirely C++ based approach to the MapReduce and its feasibility on multiple factors like developer friendliness, deployment interface, efficiency and scalability. This paper also introduces Eager Reduction and Delayed Reduction techniques that can speed up MapReduce.

I. INTRODUCTION

MapReduce is a programming paradigm for large scale data processing. Logically, each MapReduce operation consists of two phases: a map phase where each input data is mapped to a set of intermediate key/value pairs, and a reduce phase where the pairs with the same key are put together and reduced to a single key/value pair according to a user specified reduce function. This is an efficient way of parallel computing across distributed computers for big data tasks. The caveat is that the entire ecosystem revolves around JVM which was once a viable solution as it could run on any hardware and didn't run into memory and garbage allocation problems.

The original focus of Hadoop was data storage and processing which means jobs can be submitted and processed in batches which was vastly slower due to its dependence on disk. Over the last few years, the scenario has changed vastly as Big Data analytics has taken common ground which means MapReduce jobs have become more real time and the need for faster and faster outputs has become quintessential.

There are Pros and Cons of Java for using Hadoop Ecosystem [1]. Java was originally adopted for Hadoop and sequentially MapReduce to cooperate with the original Nutch framework for search engine. This method poses the following advantages:
- Detailed debugging experience.
- Mature ecosystem of developers and tools.
- Type safety and garbage collection prevent memory leaks.
- Java compiles to byte code for JVM which works on any system. Hence portable.

Java comes along with its own fair share of disadvantages like [2]:
- Memory overhead is a real problem when a JVM uses large amounts of memory just to persist which could have been used for a computational task.
- Data Skew in Hadoop's Map-reduce is a real problem when some part of workload is completed before others. [3]
- Java's implementation of data flows i.e. de-serialisation, uncompressing of records from storage is very slow due to creation and deletion of too many objects.
- Bindings is not generally possible to interface directly with Java from another language, unless that language which is used is also built on the top of the JVM. There are many problems in Hadoop that would better be solved by non-JVM language.
- Decreasing popularity of Java due to its high verbosity which prevents user adoption and rapid prototyping.
- High learning curve to get accustomed to the JVM and framework is steep.
- Iterative and incremental processing is much more difficult.
- Most of modern big data computation is dependent on memory and faster computational speeds for real time results which makes C++ sound better than Java.

Hence, it is clear that we need to explore other options that promote faster processing.

II. RELATED WORKS

There are a few other works that aim to perform the same essential task with each with their own merits and demerits.

MR-MPI [4] implementation is an Open Source C++ framework which provides C wrapper for Python. The framework works in three interface steps: Map(), collate() and

reduce() but the actual process is more like : map->aggregate->convert and then finally reduce. User provides callback functions to implement map and reduce phase while MRMPI contains functions to perform aggregate and convert. The map () is a user-defined function which generates a page memory worth of Key-Value Pairs, the pairs are communicated to other processors in an MPI_Alltoall() method. The collate () converts the KV pairs into a list of Values per key. It can also operate out-of-core which means that it can fit data out of memory by writing temporarily to disk which doesn't exceed more than 7 files. It also sorts it using Merge-Sort in O(NlogN) time based on keys. The reduce () converts the Key-Value Lists into reduced data. The framework is especially useful for graph-based algorithms. It proposes an enhanced algorithm that works well with the given framework. The graph algorithms like R-MAT, single source shortest path, maximally independent sets, triangle finding, connected components and PageRank were tested on medium-sized linux clusters and compared with MapReduce implementations. It has other useful features like in-core processing for data that fit in memory and out-of-core processing for larger datasets. The biggest disadvantage that it claims is that it isn't fault tolerant, which is caused by MPI. Current MPICH has improved the state of fault tolerance in MPI using Hydra for error reporting. The other conclusions that were drawn are that randomization of data across processors eliminates data locality but is efficient for load-balancing on even irregular data, this framework can be exploited to maintain processor specific 'state' which wouldn't have been possible in cloud-based Hadoop and that MapReduce based on MPI is easier to code and scale.

Mimir [5] is another optimization and execution framework for better MR-MPI. Mimir takes MR-MPI implementation as a base but significantly improves performance and reduces memory load using the proposed optimization techniques to reduce memory. The paper claims that MRMPI uses global barriers to synchronize data at the end of each phase which demands that all the data of the current phase has to be retained in memory or on I/O subsystem until next phase starts. This intermediate data can be very large for iterative MapReduce tasks. One of the problems of IBM like supercomputers is that it uses lightweight kernel that doesn't handle memory fragmentation due to frequent allocation and deallocation of memory buffers of different sizes. MR-MPI uses 'pages' to store intermediate data. When 'page' memory is full, it spills data via I/O to disk in an out-of-core fashion. This poses as a potential bottleneck as supercomputers don't generally have local disks and when I/O wants to write, it is written to global parallel filesystem and this makes disk-spilling expensive. MR- MPI also suffers from redundant memory buffers and unnecessary memory copies. Hence, the paper realizes a system for much more memory efficient MR-MPI implementation. It does so by introducing two more objects: called KV containers (KVCs) and KMV containers (KMVCs), to help manage MR-MPI's KVs (Key-Value pairs) and KMVs ( Key Value Merged Lists).

Blaze [6] is a similar standard for an MP/MPI based MapReduce. It is a modern C++17 based frameworks that was introduced by students of Cornell University. It brings features like Eager Reduction, Thread Local Cache and Faster Serialization. Other MPI implementations use ProtoBuf by Google to serialize and deserialize data before transmitting across processors but. It performs much better than the other libraries but has its own shortcomings. Eager reduction speeds up execution but there is no alternative for Classic MapReduce based algorithms. Eager reduction is a feature of Blaze [6] which speeds up the reduction process by reducing as soon as the map process is being done while shuffling.

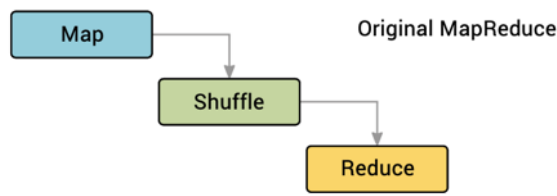

Fig. 1 Classic MapReduce pattern

Fig. 1 explains how a classic MapReduce Job is processed. First is the Map phase, where the input is passed through mappers on the cluster nodes in parallel. Then the Shuffle phase where the outputs of the map phase is transmitted across the network to the assigned Reducer. Then occurs the Reduce phase that performs the reduce operation on the accumulated results of all mappers from all the cluster nodes.

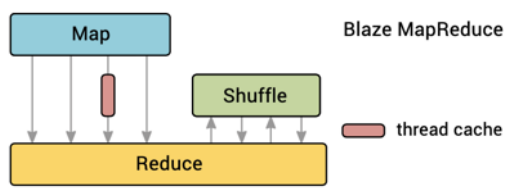

Fig. 2. Blaze MapReduce pattern with Eager Reduction

Fig. 2 explains the working of Blaze's MapReduce where the Shuffle and Reduce occur simultaneously. Reduce is applied to the output of mapper locally at the MPI slave level and then simultaneously shuffled across the network for the final shuffle phase. There is a Thread-local Cache that reduces movement of data across processors and increases execution speed by caching.

Mariane [7] is another HPC based MapReduce implementation. It is a MapReduce implementation adapted and designed to work with existing and popular distributed file systems. It can handle various cluster, shared-disk and POSIX parallel filesystems. It eliminates the need for a HDFS or other dedicated filesystem that needs to be maintained for a MapReduce system's purpose. It is said to work with NERSC, the, the Open Science Grid, NY state Grid, TeraGrid and other HPC grids based on MPI. In its implementation, it is prescribed to use Input/Output management and distribution rests within the Splitter. Concurrency management in the role of TaskController, while fault-tolerance dwells with the FaultTracker. In terms of Input Management, Mariane requires the underlying filesystem to take care of input file splitting and input distribution by utilising the 'elasticity' of cloud and other distributed storage platforms. For task management, Mariane implements a TaskTracker maintained by master node which monitors subtasks using a task completion table. If a Task failed, the FaultTracker reassigns the job based on file markers unlike Hadoop which is based on input splits. Hence achieving higher performance and speeds without compromising fault tolerance.

The paper [8] is a benchmark done to test the performance of OpenMP, MPI and Hadoop for MapReduce tasks. It misses out on one key aspect that OpenMP can be run on nodes individually on an MPI cluster which means that on each MPI node, some of the tasks can be executed in parallel along with being distributed. It states that OpenMP is easier as programmers don't have to consider workload partitioning and synchronization. It also states that "MPI allows more flexible control structures than MapReduce; hence MPI is a good choice when a program is needed to be executed in parallel and distributed manner with complicated coordination among processes".

Microsoft Azure is another cloud based platform that provides computing resources and has its proprietary MapReduce implementation too. [9]

Amazon Web Services provides Elastic Cloud Platform that implements its own Amazon Elastic File System and has its own proprietary MapReduce. [10]

Smart [11] is a MapReduce-like system for in situ data processing on supercomputing systems. Smart deviates from the traditional MapReduce to be much more adept for the needs of in situ data processing and not providing all of MapReduce semantics and interfaces.

Mimir+ [12] is a further optimized version of Mimir for GPU accelerated heterogenous systems. It was tested on Tianhe-2 supercomputer and it outperformed Mimir on data intensive tasks. They propose a pre-acceleration system that works before the GPU use and performs operations like data partitioning, data communication and data transmission.

Nguyen et. Al [13] provides an interface for using MPI on docker and using a separate registry for the containers to access files using MPI on a docker swarm.

The publication [14] provides the necessary steps needed to set up a Raspberry Pi Beowulf cluster that would be needed in to build and test the libraries discussed in the paper for small scale hardware tests.

The paper [15] provides the algorithm to implement parallel K-means algorithm using iterative Map-reduce on a distributed cluster of machines. This is later implemented in Blaze.

DELMA [16] is a proposed framework that sheds perspective on dynamic set of nodes which can be scaled up and down without interrupting the current executing jobs. The paper provides compelling reasons for a MapReduce framework to possess.

## III. PROPOSED SYSTEM

To use a HPC system as a MapReduce in production or on massively parallel systems, one of the three approaches can be adopted.

### A. Bare metal hardware

The most common configuration for most Distributed Computing Clusters is commodity hardware. Most DFS clusters run on commodity hardware. The following are installed: MPICH , OpenMP, SSH.

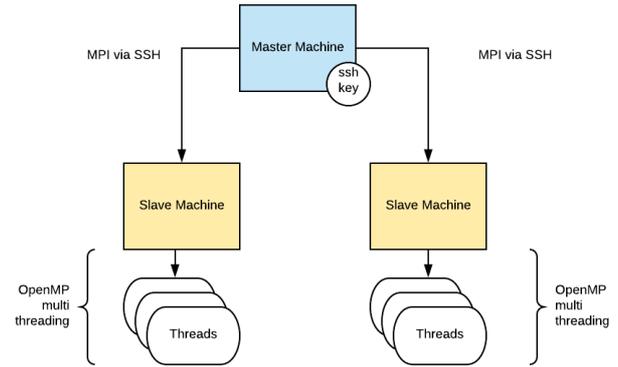

Fig. 3. Architecture for Bare metal MPI cluster with OpenMP

Fig. 3 describes a common setup for HPC applications where Master and Slave nodes communicate via MPI using OpenSSH. Each node also processes in parallel using OpenMP individually. SSH key exchange and password-less SSH is required for the communication to take place.

### B. VM cluster based MapReduce

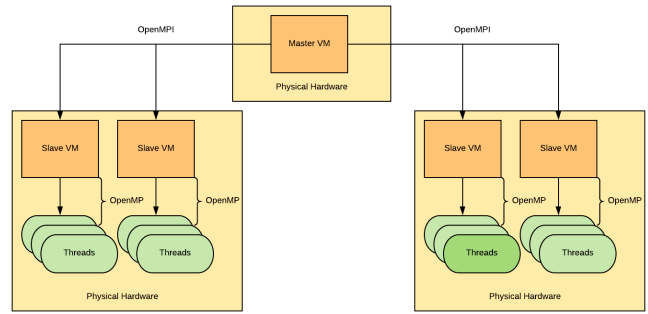

Fig. 4. Architecture for VM based MPI cluster with OpenMP

Fig. 4 explains a better and more efficient setup for clusters by using VMs. This is especially advantageous as it reduces the cost of setting up and reduces hassle. It offers other advantages such as environment isolation , easier recovery and faster updates to environment and easier maintenance . The only disadvantage is the overhead imposed by hypervisor which causes increased boot up times and slower performance on some instructions.

### C. Containerization based MapReduce

The paper proposes a Docker or Singularity based containerized application. The advantages of containerizing is many. The first and foremost important aspect is portability. Apart from portability, they provide efficient utilization of hardware and with orchestration methods such as Kubernetes, it becomes easier to recover from failure and maintain ready resources and application states.

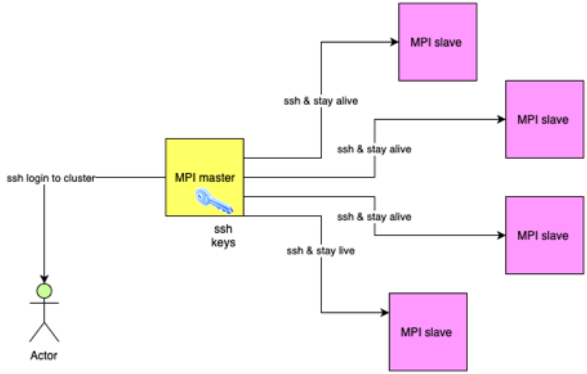

Fig. 5. Architecture for Containers based MPI cluster with OpenMP

Fig 5. Explains a container based MPI clusters where a Docker service for SSH has been separately deployed and used for communication services. The MPI slaves are containers created from MPI slave images while the master is from the master image. During the automated setup using docker-swarm utility, the SSH keys are copied from master image to slave image and an endpoint is provided for the user to communicate to the master container. This type of MPI cluster setup is massively advantageous as it provides fault tolerance, faster setup, better utilization of resources, portability and efficient updates to software. In contrast to the VMs, containerized approach has negligible overhead.

*D. Delayed Reduction*

When the framework was used to develop other algorithms like matrix multiplication and linear regression, it felt rigidity due to the eager reduction and it was almost impossible to implement the algorithms.

This becomes an issue when reduction has to be done over the iterable list of HashMap which currently is not possible in Blaze framework. Hence, to solve this issue, Delayed Reduction has been added to framework as shown in Fig. 6.

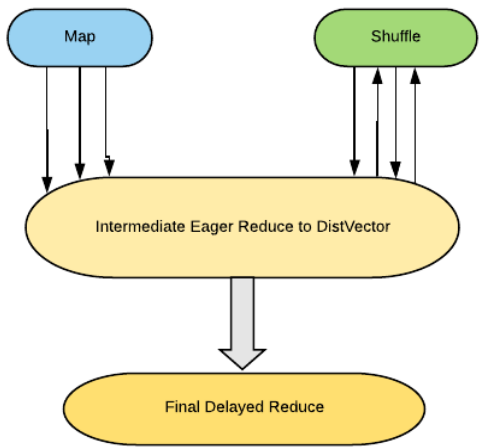

Fig. 6. Improved Blaze MapReduce pattern with Delayed Reduction

To implement Delayed Reduction, we need to understand two APIs provided by Blaze: DistVector and DistHashmap.

・A DistHashmap works by sharding, balancing and distributing a standard C++ based HashMap over an MPI cluster using collective communication.
・The DistVector is based on DistHashMap but with serial keys that converts a C++ standard vector into a DistHashMap and then shards, balances and distributed across the MPI cluster.

In eager reduction, the reduction occurs as soon as part of mapper process completes between two values with the same key. As an alternative to this Delayed Reduction works by creating a temporary DistVector and emitting as the output of a mapper which contains all the locally reduced values. This DistVector is reduced immediately into another DistVector after sorting using Merge Sort and then shuffled across the network. The final reducer does the actual reduction job but now it operates on an Iterable of Values instead of a single value, as would have been possible with Eager Reduction.

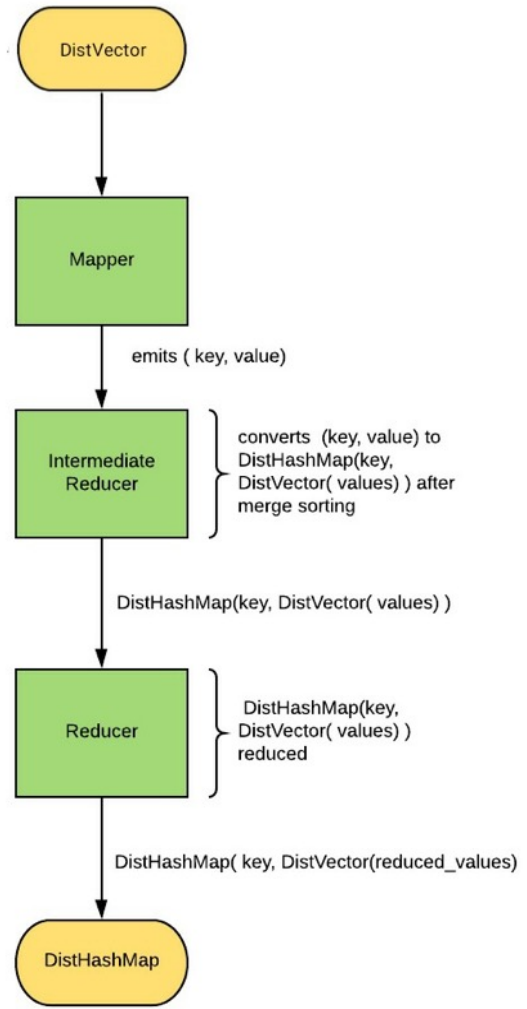

Fig. 7. Algorithm for Delayed Reduction

The Fig. 7 can be more formally represented by the following pseudocode.

*Pseudocode:*

1) A DistVector or DistHashMap or a C++ STL vector contains the source for MapReduce.
2) Mapper can be any function that emits a ( Key, Value) pair and acts on the Source.
3) Intermediate reducer combines the keys into a DistVector.
4) MapReduce is called on the source DistVector to convert it into a ( Key, Iterable<Value> ). This DistVector is distributed across the cluster in-memory.
5) The final Reducer works on an Iterable of Values now. This can be called immediately or later. Laziness of Reduction is displayed.
6) The final DistHashMap is used to hold final Reduced HashMap in a distributed manner.

## IV. EXPERIMENTAL SETUP

There are three ways an HPC system based MapReduce framework can be used in increasing scale:
- Virtual machines that are configured to run on hardware.
- Directly on commodity grade hardware.
- Containerized images using Docker or Singularity.

The steps to setup and configure a working cluster on the three different hardware is as follows.

### A. Raspberry pi MPI cluster

The experimental setup that was used to simulate commodity hardware was Raspberry pi 3B+ with 1GB LPDDR2 SDRAM, Gigabit Ethernet, MPICH2. Steps to setup an MPI cluster on an array of Raspberry Pi are as follows:

Install Raspbian OS and install physical ethernet network. Dedicate one RPI as master and the rest as slaves. Install MPICH2 on all devices. Enable password-less SSH from master to all the slaves. Create Hostfile with all the IP addresses of the slaves. Mpirun the code with the library in path , along with hostfile each time.

### B. VM MPI cluster

The Ubuntu 18.04 VMs with configurations of 4 GB RAM , 10 GB SSD , VirtualBox images on Ubuntu 18.04, OpenMPI 2.1.1 are used. Steps to setup an MPI cluster on an array of VirtualBox VMs:

Create an Ubuntu 18.04 VM and allocate resources and assign a static IP on a Bridge Network. Install OpenMPI and accompanying libraries on the created VM. This VM will be master node. Make clones of the created VM for slave nodes. Update to unique IPs of the slave VMs. Enable password-less SSH from master to all the slaves. Create Hostfile with all the IP addresses of the slaves. Mpirun along with hostfile each time.

### C. Containerized application using Docker

The steps to set up containerized is as follows using the docker image from [17] :

Install Docker daemon on host using  *$ sudo apt-get install docker.* Pull nlknguyen/alpine-mpich from Docker hub  using the command *$ docker pull nlknguyen/alpine-mpich* . Use this command to run an MPI image for development, $ *docker run --allow-run-as-root --rm -it -v '$(pwd):/project' nlknguyen/alpine-mpich*. Use the existing Docker-swarm configuration to create an MPI cluster.  *$  ./cluster/cluster.sh up size=2*. Use this command to execute any MPI run command.  *$ ./cluster/cluster.sh exec <command>*.

## V. IMPLEMENTATION AND RESULTS

The library of choice to establish a complete MapReduce system was Blaze [18] a modern alternative to Google MR-MPI with features like eager reduction, thread local cache and fast serialization which has the potential to boost performance of MapReduce algorithms. The problem caused by the lack of fault tolerability still persists due to the MPI implementation.

A custom test was devised to measure on the following aspects on real hardware for testing performances in smaller key ranges and datasets for performance, scalability and adaptability for different algorithms.

For larger datasets, it is compared against the Spark implementation of the algorithms using MLlib library that implements machine learning algorithms.

### A. K-Means Clustering using MapReduce

K-means clustering is an unsupervised machine learning algorithm for classification. This can be implemented in MapReduce using the algorithm described in [15].

The following results were observed when the framework was used for K-means clustering program :

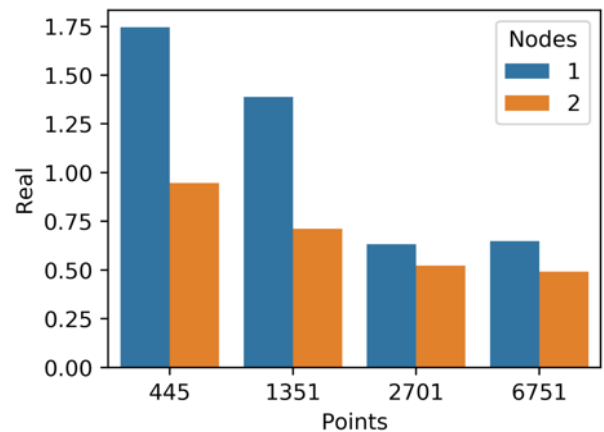

Fig. 8. K-means Clustering on Blaze framework

As seen from Fig 8, K-means performance was optimal and with increasing dimensions, the algorithm performed better. Scalability was displayed with increasing performance with nodes.

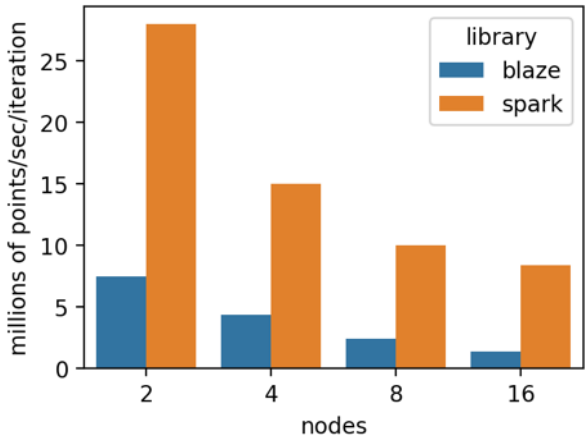

Fig. 9. K-means Clustering comparison between Blaze and Spark

As seen in Fig 9, K - Means clustering on Blaze was tested to be faster than Spark implementation by a large margin. The scalability was close to linear and halved for each rise in number of nodes.

### B. Word Count using MapReduce

The time taken for each is measured against number of points and number of nodes processing. It is observable that the framework tends to increase performance as number of nodes increases and displays linear scalability as observable in Fig 10.

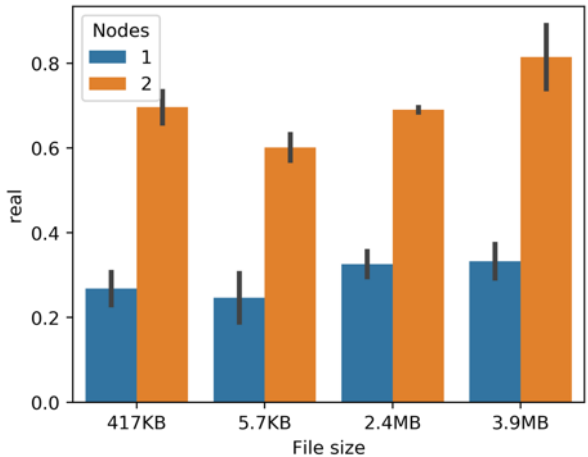

Fig. 10. Wordcount on VM cluster using Blaze framework

WordCount is the hello-world program of MapReduce. This task was inefficient in terms of scalability as the framework tended to increase processing time with increase in nodes. This has to be fixed in future work. Part of issue of scalability can be addressed to the shuffle phase unable to facilitate movement of large loads of KV pairs which is unsuitable for low key ranges but on larger dataset, the scalability is linear as seen in Fig.11.

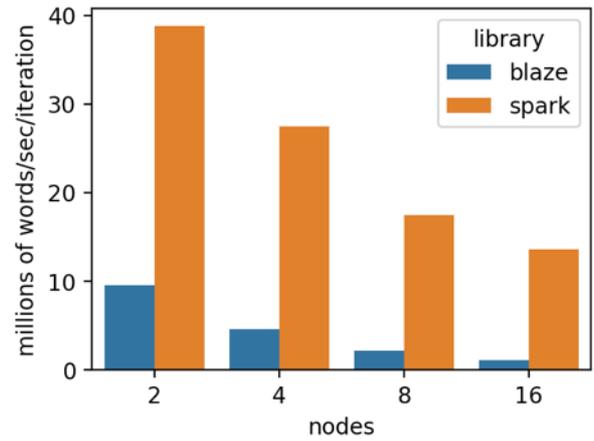

Fig. 11. Wordcount comparison between Blaze and Spark

### C. Pi Estimation using MapReduce

Pi estimation using Monte Carlo is an algorithm where random coordinates (x,y) are generated in mappers and if they fall within a certain range the mapper emits (key,1), else emits (key, 0). The reducer sums over the key and estimates the value of pi using 4 * (count of points inside/ total count of points).

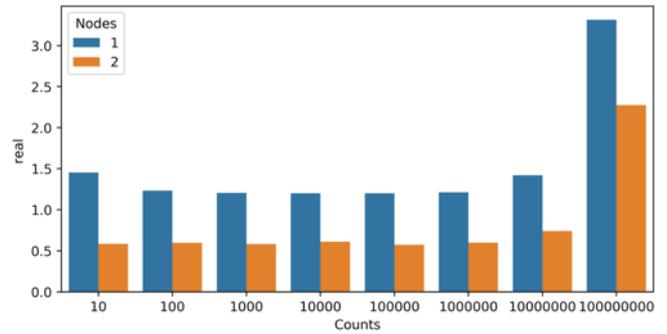

Fig. 12. Pi estimation using Monte Carlo method on VM cluster using Blaze framework

As seen in Fig. 12, this algorithm when implemented on the framework, was very efficient in terms of memory, speed and scalability. The time taken for processing reduces almost linearly for increase in number of nodes.

### D. Peak Memory Usage Comparison

The overall Memory usage was compared for different algorithms against Blaze and Spark to find differences in performance and efficiency and plotted in Fig 13.

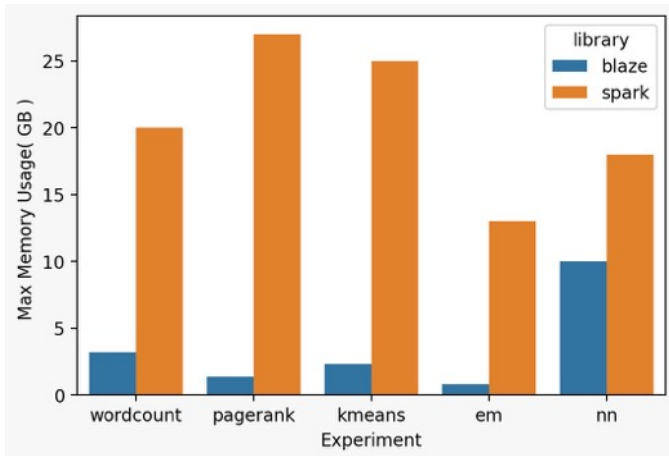

Fig. 13. Memory Usage difference between Blaze framework and Spark

VI.  CONCLUSIONS AND FUTURE WORK

An entirely C++ based framework for MapReduce is very efficient and fast and compares well with the standard implementation. From a developer standpoint, it provided a simple and code efficient implementation. From deployment standpoint, the MPI isn't fault tolerant being one of the bottleneck to the proposed system. The proposed architecture that is provided with this paper can provide a feasible and alternative to MapReduce based on JVM. This paper provides an additional feature for Delayed Reduction to provide optimized and fully featured MapReduce that is similar to Hadoop's implementation in terms of features but faster and easier to develop and deploy as a HPC based alternative to JVM based Hadoop MapReduce.

This paper implements a few MapReduce based machine learning algorithms on the given framework but a lot more algorithms can be implemented and tested in the future to prove the efficiency that HPC based systems are far faster on compute intensive tasks than in JVM based systems.

VII.  ACKNOWLEDGEMENT

I would like to thank Vellore Institute of Technology, Chennai for providing the opportunity and support to work on this paper.

IX.  Authors

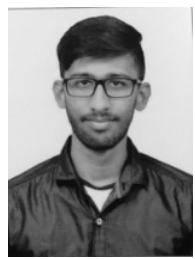

Vignesh S., is a B.Tech undergraduate student of Computer Science Engineering from Vellore Institute of Technology, Chennai campus. His main fields of interests are Big Data, Machine Learning, Parallel and Distributed Computing.
Email:
**s.vignesh2017a@vitstudent.ac.in**

Dr. V. Muthumanikandan M.E., Ph.D. is working as a Senior Assistant Professor in the School of Computer Science and Engineering, VIT University, Chennai, India. He received his

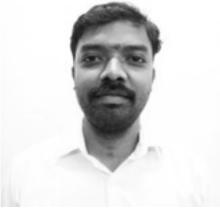
B.E and M.E degree in Computer Science and Engineering discipline. He published many papers in the conferences and reputed journals. His areas of research interests include Parallel and Distributed Systems, Networking, Software Defined Networking and Network Function Virtualization.
Email: **muthumanikandan.v@vit.ac.in**

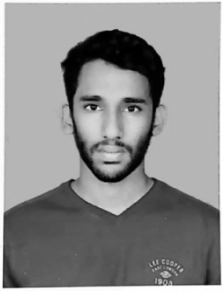
Siddarth Sairaj., is a B.Tech undergraduate student of Computer Science Engineering from Vellore Institute of Technology, Chennai campus. His main fields of interests are Machine Learning, High Performance Computing and Distributed Systems.
Email:
**siddarthsairaj2017@vitstudent.ac.in**

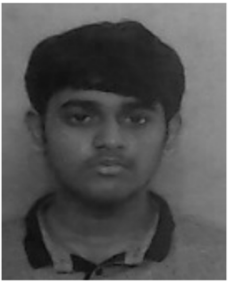
Sainath Ganesh., is a B.Tech undergraduate student of Computer Science Engineering from Vellore Institute of Technology, Chennai campus. His main fields of interests are evolutionary algorithms in Artificial Intelligence, Augmented Reality and Distributed Systems.
Email:
**sainath.g2017@vitstudent.ac.in**